\documentclass{nature}
\usepackage{amsmath,amssymb,graphicx,hyperref,url}  
\usepackage[none]{hyphenat}
\usepackage[usenames]{color}

\newcounter{firstbib}
\newcommand{\tc}{\textcolor{black}}

\def\etal{{\frenchspacing\it et al.}}

\def\eg{{\frenchspacing\it e.g.}}

\def\be{\begin{equation}}
\def\ee{\end{equation}}
\def\ba{\begin{eqnarray}}
\def\ea{\end{eqnarray}}

\title{\tc{Dynamical dark energy in light of the latest observations}}

\author{Gong-Bo Zhao$^{1,2}$, 
Marco Raveri$^{3,4}$, 
Levon Pogosian$^{5,2}$,
Yuting Wang$^{1,2}$,
Robert G. Crittenden$^{2}$,
Will J. Handley$^{6,7}$,
Will J. Percival$^{2}$,
Florian Beutler$^{2}$,
Jonathan Brinkmann$^{8}$,
Chia-Hsun Chuang$^{9,10}$,
Antonio J. Cuesta$^{11,12}$,
Daniel J. Eisenstein$^{13}$,
Francisco-Shu Kitaura$^{14,15}$,
Kazuya Koyama$^{2}$,
Benjamin L'Huillier$^{16}$,
Robert C. Nichol$^{2}$,
Matthew M. Pieri$^{17}$,
Sergio Rodriguez-Torres$^{9,18,19}$,
Ashley J. Ross$^{20,2}$,
Graziano Rossi$^{21}$,
Ariel G. S\'anchez$^{22}$,
Arman Shafieloo$^{16,23}$,
Jeremy L. Tinker$^{24}$,
Rita Tojeiro$^{25}$,
Jose A. Vazquez$^{26}$
\& Hanyu Zhang$^{1}$
}

\begin{document}
\maketitle

\begin{abstract}

A flat Friedman-Roberson-Walker universe dominated by a cosmological constant ($\Lambda$) and cold dark matter (CDM) has been the working model preferred by cosmologists since the discovery of cosmic acceleration\cite{Riess98,Perl99}. However, tensions of various degrees of significance are known to be present among existing datasets within the $\Lambda$CDM framework\cite{Delubac14,Sahni:2014ooa,Battye14,Aubourg14,P15,Raveri15,Addison15,Bernal16,Freedman17}. In particular, the Lyman-$\alpha$ forest measurement of the Baryon Acoustic Oscillations (BAO) by the Baryon Oscillation Spectroscopic Survey (BOSS)\cite{Delubac14} prefers a smaller value of the matter density fraction $\Omega_{\rm M}$ compared to the value preferred by cosmic microwave background (CMB). Also, the recently measured value of the Hubble constant, $H_0=73.24\pm1.74 \ {\rm km}\ {\rm s}^{-1} \ {\rm Mpc}^{-1}$\cite{Riess16}, is $3.4\sigma$ higher than $66.93\pm0.62 \ {\rm km}\ {\rm s}^{-1} \ {\rm Mpc}^{-1}$ inferred from the Planck CMB data\cite{P15}. \tc{In this work, we investigate if these tensions can be interpreted as evidence for a non-constant dynamical dark energy (DE). Using the Kullback-Leibler (KL) divergence\cite{Kullback:1951va} to quantify the tension between datasets, we find that the tensions are relieved by an evolving DE, with the dynamical DE model preferred at a $3.5\sigma$ significance level based on the improvement in the fit alone.} While, at present, the Bayesian evidence for the dynamical DE is insufficient to favour it over $\Lambda$CDM, we show that, if the current best fit DE happened to be the true model, it would be decisively detected by the upcoming DESI survey\cite{DESI}.
\end{abstract}

The observational datasets considered in this work include the latest CMB temperature and polarisation anisotropy spectra, the supernovae (SNe) luminosity distance data, the BAO angular diameter distance data from the clustering of galaxies (gBAO) and from the Lyman-$\alpha$ forest (Ly$\alpha$FB), the measurement of $H_0$, $H(z)$ measurements using the relative age of old and passively evolving galaxies (OHD), the three-dimensional galaxy power spectra, and the two-dimensional weak lensing shear angular power spectra. \tc{Further details about the datasets and associated systematic effects can be found in {\bf Methods}.}

The KL divergence, also known as relative entropy, quantifies the proximity of two probability density functions (PDFs). Rather than focusing on particular model parameters, it is designed to compare the overall concordance of datasets within a given model. We use the difference between the actual and the expected KL divergence, called ``Surprise''\cite{Seehars:2014ora}, as a measure of tension between datasets. Rather than comparing the PDFs for the $\Lambda$CDM parameters for every pair of datasets, we take the combined dataset, ALL16 (see Supplementary Table 1), and find the derived PDFs for the angular diameter distance $D_A(z)$ and the Hubble parameter $H(z)$ at redshifts corresponding to the available data. We then compute the KL divergence between the derived PDFs and the directly observed $D_A(z)$ and $H(z)$ from $H_0$, SNe, OHD, gBAO and Ly$\alpha$FB, and evaluate the corresponding Surprise and the standard deviation (see {\bf Methods} for details). Results are shown with \tc{cyan} bars in Fig.~1\tc{a}. They indicate that the $H_0$, Ly$\alpha$FB and SNe measurements are in tension with the combined dataset. Introducing Tension $T$ as the number of standard deviations by which Surprise is greater than zero, we find values of $T=4.4$, $3.5$, and $1.7$ for the $H_0$, Ly$\alpha$FB and SNe measurements, respectively \tc{\bf(shown in Fig.~1b)}, with the first two values signalling significant tension.

Next, we check if the tension within the $\Lambda$CDM model can be interpreted as evidence for a dynamical DE. The dynamics of DE can be probed in terms of its equation of state $w$, which is equal to $-1$ for $\Lambda$, \tc{but is different in dynamical DE models where it will generally be a function of redshift $z$. Commonly considered alternatives to $\Lambda$ are a model with a constant $w$ ($w$CDM), and one in which $w$ is linear function of the scale factor ($w_0w_a$CDM)\cite{CPL}}. We allow for a general evolution of the DE equation of state and use the correlated prior method\cite{Crittenden11} to perform a Bayesian non-parametric reconstruction of $w(z)$ \tc{using the Monte Carlo Markov Chain method with other cosmological parameters marginalised over (see {\bf Methods} for details)}. Fig.~2 presents the reconstructed $w(z)$, along with the 68\% confidence level (CL) uncertainty, shown with a light blue band, derived from the combined dataset ALL16. Table 1\tc{a} shows the change in $\chi^2$ relative to $\Lambda$CDM for each individual dataset for the best fit $w(z)$CDM model derived from ALL16. Overall, the $\chi^2$  is improved by $-12.3$, which can be interpreted as the reconstructed dynamical DE model being preferred at $3.5\sigma$. The reconstructed DE equation of state evolves with time and crosses the $-1$ boundary, which is prohibited in single field minimally coupled quintessence models\cite{Vikman04}, \tc{but can be realised in models with multiple scalar fields, such as Quintom\cite{quintom}, or if the DE field mediates a new force between matter particles\cite{Das05}. In the latter case, which is commonly classified as Modified Gravity, it is quite generic for the effective DE equation of state to be close to $-1$ around $z=0$, but evolve towards more negative values at intermediate redshifts, before eventually approaching $0$ during matter domination. Such dynamics would be consistent with our reconstruction and could be tested in the future when BAO measurements at higher redshifts become available.} In addition to the reconstruction from the combined ALL16 dataset presented in Fig.~2, we present reconstructions derived from ten different data combinations in Supplementary Fig.~1. 

The results for tension between datasets, re-evaluated for the ALL 16 best fit $w(z)$CDM model, are shown with dark blue bars in Fig.~1. We find $T=0.7$, $1.1$ and $0.7$ for $H_0$, Ly$\alpha$FB and SNe, respectively, indicating that tensions that existed in the $\Lambda$CDM model are significantly released within $w(z)$CDM.  A plot of the relevant data points along with the best fit predictions from the $\Lambda$CDM and the $w(z)$CDM model are provided in the Supplementary Fig.~2. 

With a large number of additional $w$-bin parameters, one may be concerned that the improvement in the fit is achieved by $w(z)$CDM at the cost of a huge increase of the parameter space. However, correlations between the $w$-bins induced by the prior constrain most of that freedom. One way to estimate the effective number of additional degrees of freedom is to perform a principal component analysis (PCA) of the posterior covariance matrix of the $w$-bin parameters and compare it to that of the prior. Using this method, explained in detail in {\bf Methods}, we find that our $w(z)$CDM model effectively has only four additional degrees of freedom compared to $\Lambda$CDM. We note that the demonstration that ALL16 is capable of constraining four principal components of $w(z)$ is one of the interesting results of this work.

It is interesting to compare $w(z)$ reconstructed from ALL16 to that obtained in Zhao \etal~(2012)\cite{Zhao12} using the same prior but a different dataset which we call ALL12 (a comparison of the ALL16 and the ALL12 datasets is provided in Supplementary Table 2). ALL16 contains about 40\% \tc{new} supernovae \tc{compared to} ALL12, primarily provided by the SDSS-II survey. Moreover, in ALL12, the BAO measurement derived from the BOSS DR9 sample\cite{dr9} was at a single effective redshift, while in ALL16 it is tomographic at nine redshifts from BOSS DR12\cite{Zhao17}, which contains four times more galaxies than DR9. In addition, ALL16 includes a high-redshift BAO measurement from Lyman-$\alpha$ forest, \tc{which was not available in 2012.} This \tc{helps} to constrain $w(z)$ at redshifts where \tc{the supernovae constraints are weak}. The new 2016 $H_0$ measurement\cite{Riess16} is consistent with that in 2009\cite{Riess09}, with the error bar halved. \tc{Comparing measurements of the expansion rate and the cosmic distances in ALL12 and ALL16, we find that those in ALL16 offer information at more redshift values, and with a greater signal-to-noise (S/N) ratio (see Supplementary Figure 4 for a visual comparison). Quantitatively, the S/N in measurements of $H(z)$, $D_A(z)$ and $d_L(z)$ in ALL16 is larger by 80\%, 260\% and 90\%, respectively, compared to the ALL12 dataset.} The Planck 2015 CMB data is also much more informative than the WMAP 7-year release\cite{wmap7}, thanks to a higher angular resolution of the temperature and polarisation maps, and lower levels of statistical uncertainties.  

Overall, ALL16 is more constraining \tc{due to a significant level of new and independent information in ALL16 compared to ALL12}: the effective number of $w(z)$ degrees of freedom constrained by ALL12 was three, compared to four constrained by ALL16. Fig. 2 compares the two results and shows that they are highly consistent. We quantify the agreement by evaluating the dot-product of the $\hat{\bf w}$ vectors from the two reconstructions (the vectors are normalised so that a dot-product is unity if they are identical) \tc{and} find that $\hat{\bf w}_{\rm ALL12}\cdot\hat{\bf w}_{\rm ALL16}=0.94\pm0.02.$ We also evaluate the \tc{tension $T$} between the two \tc{reconstruction results} and find that $T=-1.1$. \tc{This indicates an excellent alignment of the two results. This agreement, and the raised significance of an evolving $w(z)$ from $2.5\sigma$ to $3.5\sigma$ CL with more advanced observations, suggests the possibility of revealing the dynamics of DE at a much more statistically significant level in the near future, as we will present later.} 

To check \tc{whether} the improvement in the fit warrants introducing additional effective degrees of freedom, we evaluate and compare the Bayesian evidence, $E \equiv \int d \theta \, {\cal L}({\bf D}|\theta)\, P(\theta)$, for $\Lambda$CDM and the $w(z)$CDM model. The Bayes' factors, which are the differences in $\ln E$ between the two models, are shown \tc{in} Table~1. The Bayes' factors for both \tc{the ALL12 and the ALL16 DE models} are negative, indicating that $\Lambda$CDM is favoured by this criterion. However, our forecast for a future dataset, DESI++, comprised of BAO measurements from DESI\cite{DESI}, \tc{around 4000 supernovae luminosity distances} from future surveys\cite{EuclidSN} and CMB (assuming the Planck sensitivity), predicts a Bayes' factor of $11.3 \pm 0.3$ if the ALL16 $w(z)$CDM reconstruction happened to be the true model, which would be decisive according to the Jeffreys scale.

One may ask how much the evidence for DE depends on the particular choice of the prior parameters. In principle, the choice of the smoothing scale should be guided by theory. The value used in Zhao \etal~(2012)\cite{Zhao12} and this work, $a_c=0.06$, \tc{is a time-scale conservatively chosen to be sufficiently small not to bias reconstructions of $w(z)$ expected in quintessence DE models\cite{Crittenden11}.} \tc{For} the inference to be conclusive, the evidence for a dynamical DE should be strong over a wide range of the prior parameters. \tc{Therefore, we vary the strength of our prior by adjusting $\sigma_{\rm D}$, a parameter added to the diagonal of the inverse of the prior covariance matrix, and examine how the significance of the dynamical DE detection, as well as the Bayesian evidence, change with the variation of $\sigma_{\rm D}$. As shown in Fig.~3, and with additional details given in {\bf Methods},} we find that neither ALL12 nor ALL16 provide evidence for a dynamical DE over the considered wide range of prior strengths. \tc{However, the Bayes factor for ALL16 is generally much less negative than that of ALL12 for all prior strengths, \eg, it increased from $-6.7\pm0.3$ to $-3.3\pm0.3$ for $\sigma_{\rm D}=3$, which is the prior used in this work. In fact, for ALL16, the Bayes factor remains close to zero for $\sigma_{\rm D}\lesssim0.4$. We plot the model with $\sigma_{\rm D}=0.4$ as a light green band in Fig.~2 to demonstrate the impact of changing the prior strength, and also because it is a model that has the same Bayesian evidence as $\Lambda$CDM while deviating from $\Lambda$ at a $2.7\sigma$ CL.} On the other hand, our forecast for DESI++ shows that, if the $w(z)$CDM model was true, it would be decisively supported by Bayesian evidence over a wide range of prior strengths, \tc{as shown in Fig.~3.}

\tc{Various ways to relieve the tension between datasets have been proposed, including allowing for additional relativistic degrees of freedom\cite{DiValentino16}, massive neutrinos\cite{DiValentino16}, and interacting vacuum\cite{Wang15,Sola16}. In addition, to relieve the tension between the  $\Lambda$CDM parameters required to fit the CMB temperature anisotropy spectrum at large and small scales, the Planck team introduced\cite{P15} a parameter $A_{\rm Lens}$ that rescales the amplitude of the weak lensing contribution to the temperature power spectrum. In the $w(z)$ reconstruction discussed earlier, we fixed $A_{\rm Lens}=1$, assumed that neutrinos are massless and set the effective number of relativistic species at the standard $\Lambda$CDM value of $N_{\rm eff}=3.04$. We have checked the effect of these parameters on the reconstructed $w(z)$ by considering model $M_1$, with $M_{\nu}$ fixed to $0.06$ eV, model $M_2$ , with $M_{\nu}$ and $N_{\rm eff}$ added as free parameters, and model $M_3$, with varied $A_{\rm Lens}$. In all these cases, we find that the shape of the reconstructed $w(z)$ and the significance of its deviation from $-1$ are practically the same. The inferred values of the cosmological parameters in these models are given in Supplementary Table 3, and the corresponding reconstructed $w(z)$ are shown in Supplementary Fig.~5. The Bayes factors for $M_1$, $M_2$ and $M_3$ relative to the corresponding $\Lambda$CDM models (with the same added parameters), are shown in Table 1c. We also checked that neither the constant $w$ model ($M_4$), nor the linear ($w_0,w_a$) parametrisation of $w$ ($M_5$), are capable of releasing the tensions between all datasets simultaneously (see Table 1a). Interestingly, we find that our DE model with a non-parametrically reconstructed  $w(z)$ has a larger Bayes factor compared to $w_0w_a$CDM despite the latter having only two parameters (see results for $M_4$ and $M_5$ in Table 1c).}

\tc{There is always a possibility that the tensions between datasets, quantified in terms of the KL divergence in this work, are due to yet unknown systematic effects. However, it is intriguing that they persist with improvements in the quantity and the quality of the data\cite{Bernal16,Freedman17,DR12LyA}. If interpreted as a manifestation of DE dynamics, they suggest a $w(z)$ that crosses $-1$ and has a shape that is representative of modified gravity models. The commonly used ($w_0,w_a$) parametrisation would have missed this behaviour and has a lower Bayesian evidence than the reconstructed $w(z)$ model, despite the latter having more degrees of freedom. Thus, our results demonstrate that the current data can provide non-trivial constraints on the DE dynamics. It is also intriguing that the evidence for $w(z) \ne -1$, while below that of $\Lambda$CDM, has become stronger with the new independent data added since 2012, and that the ALL16 reconstruction is consistent with the ALL12 $w(z)$. We emphasise that we have not optimised the prior to maximise either the Bayes ratio or the statistical significance of the departure from $-1$, as it would be contrary to the principles of Bayesian inference. Future data has the ability to conclusively confirm the DE evolution inferred in this work if it happened to be the one chosen by Nature.}

\begin{addendum}

 \item[Correspondence] Correspondence and requests for materials should be addressed to G.B.Zhao \\ (email: gbzhao@nao.cas.cn).

 \item[Acknowledgements] G.B.Z. is supported by NSFC Grant No. 11673025, and by a Royal Society-Newton Advanced Fellowship. G.B.Z. and Y.W. are supported by National Astronomical Observatories, Chinese Academy of Sciences and by University of Portsmouth. M.R. is supported by U.S. Dept. of Energy contract DE-FG02-13ER41958. M.R. acknowledges partial support, during the development of this work, by the Italian Space Agency through the ASI contracts Euclid-IC (I/031/10/0) and the INFN-INDARK initiative. M.R. thanks SISSA where part of this work was completed. L.P. is supported by NSERC, RC by STFC grant ST/H002774/1, and Y.W. by NSFC grant No. 11403034. G.R. acknowledges support from the National Research Foundation of Korea (NRF) through NRF-SGER 2014055950 funded by the Korean Ministry of Education, Science and Technology (MoEST), and from the faculty research fund of Sejong University in 2016. A.S. would like to acknowledge the support of the National Research Foundation of Korea (NRF - 2016R1C1B2016478).

Funding for SDSS-III has been provided by the Alfred
P. Sloan Foundation, the Participating Institutions, the
National Science Foundation, and the U.S. Department
of Energy Office of Science. The SDSS-III web site is
\url{http://www.sdss3.org/}. SDSS-III is managed by the Astrophysical Research
Consortium for the Participating Institutions of the SDSS-III Collaboration including the University of Arizona, the
Brazilian Participation Group, Brookhaven National Laboratory, Carnegie Mellon University, University of Florida,
the French Participation Group, the German Participation Group, Harvard University, the Instituto de Astrosica de Canarias, the Michigan State/Notre Dame/JINA Participation Group, Johns Hopkins University, Lawrence Berkeley National Laboratory, Max Planck Institute for Astrophysics, Max Planck Institute for Extraterrestrial Physics,
New Mexico State University, New York University, Ohio
State University, Pennsylvania State University, University
of Portsmouth, Princeton University, the Spanish Participation Group, University of Tokyo, University of Utah,
Vanderbilt University, University of Virginia, University of
Washington, and Yale University.

 \item[Author contributions] G.B.Z. proposed the idea, performed the dark energy reconstruction, evidence calculation, principal component analysis and the tension calculation. M.R. and Y.W. contributed to the tension calculation. G.B.Z. and L.P. wrote the draft, and all other co-authors commented on and helped improving the manuscript, and/or contributed to the BOSS data analysis. 
 
 \item[Author Information] 
 \begin{affiliations}
 \item National Astronomy Observatories,
Chinese Academy of Science, Beijing, 100012, P.R.China
 \item Institute of Cosmology and Gravitation, University of Portsmouth,
Portsmouth, PO1 3FX, UK
 \item Kavli Institute for Cosmological Physics, Enrico Fermi Institute, The University of Chicago, Chicago, Illinois 60637, USA
 \item Institute Lorentz, Leiden University, PO Box 9506, Leiden 2300 RA, The Netherlands
 \item Department of Physics, Simon Fraser University, Burnaby, BC, V5A 1S6, Canada
 \item Astrophysics Group, Cavendish Laboratory, J. J. Thomson Avenue, Cambridge, CB3 0HE, UK
 \item Kavli Institute for Cosmology, Madingley Road, Cambridge, CB3 0HA, UK
 \item Apache Point Observatory, P.O. Box 59, Sunspot, NM 88349, USA
 \item Instituto de F\'{\i}sica Te\'orica, (UAM/CSIC), Universidad Aut\'onoma de Madrid,  Cantoblanco, E-28049 Madrid, Spain
 \item Leibniz-Institut f\"ur Astrophysik Potsdam (AIP), An der Sternwarte 16, 14482 Potsdam, Germany
 \item Institut de Ci\`encies del Cosmos (ICCUB), Universitat de Barcelona (IEEC- UB), Mart\'{\i} i Franqu\`es 1, E-08028 Barcelona, Spain
 \item Departamento de F{\'\i}sica, Universidad de C{\'o}rdoba, Campus de Rabanales, Edificio Albert Einstein, E-14071 C{\'o}rdoba, Spain
 \item Harvard-Smithsonian Center for Astrophysics, 60 Garden St., Cambridge, MA 02138, USA
 \item Instituto de Astrof isica de Canarias, 38205 San Crist obal de La Laguna, Santa Cruz de Tenerife, Spain
 \item Departamento de Astrof i sica, Universidad de La Laguna (ULL), E-38206 La Laguna, Tenerife, Spain
 \item Korea Astronomy and Space Science Institute, 776 Daedeokdae-ro, Yuseong-gu, Daejeon 34055, Korea
 \item Aix Marseille Univ, CNRS, LAM, Laboratoire d'Astrophysique de Marseille, Marseille, France
 \item Campus of International Excellence UAM+CSIC, Cantoblanco, E-28049 Madrid, Spain
 \item Departamento de F\'{\i}sica Te\'orica, Universidad Aut\'onoma de Madrid, Cantoblanco, E-28049, Madrid, Spain
 \item Center for Cosmology and AstroParticle Physics, The Ohio State University, Columbus, OH 43210, USA
 \item Department of Astronomy and Space Science, Sejong University, Seoul 143-747, Korea
 \item Max-Planck-Institut f\"ur extraterrestrische Physik, Postfach 1312, Giessenbachstr., 85741 Garching, Germany
 \item University of Science and Technology, 217 Gajeong-ro, Yuseong-gu, Daejeon 34113, Korea
 \item Center for Cosmology and Particle Physics, Department of Physics, New York University, 4 Washington Place, New York, NY 10003, USA
 \item School of Physics and Astronomy, University of St Andrews, North Haugh, St Andrews, KY16 9SS, UK
 \item Brookhaven National Laboratory, Bldg 510, Upton, New York 11973, USA
 
\end{affiliations}

 \item[Competing Interests] The authors declare that they have no
competing financial interests.
\end{addendum}

\clearpage

\begin{figure}
\begin{center}
\includegraphics[width=0.99\linewidth]{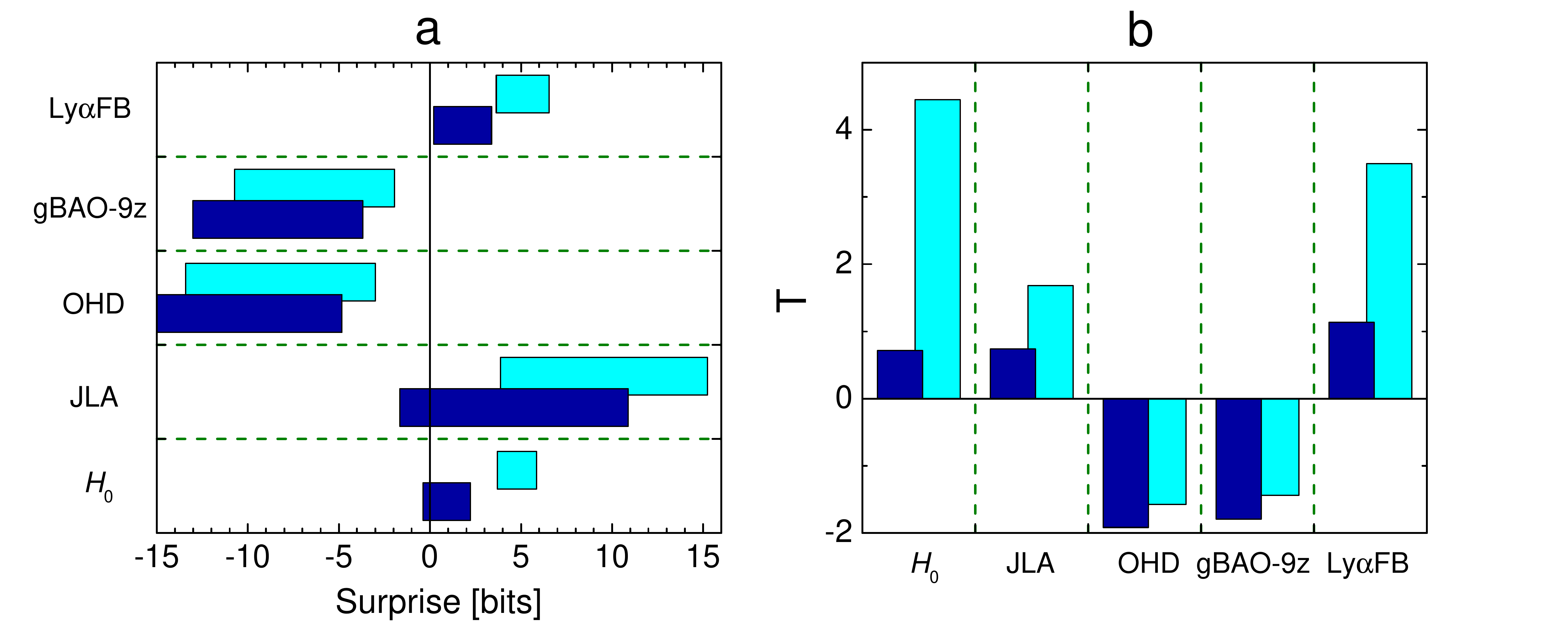}
\end{center}
\end{figure} 

\noindent {\bf Figure 1. The tension among different datasets in $\Lambda$CDM and $w(z)$CDM universes.}\\
Panel a: The Surprise between the PDFs for $D_A(z)$ and $H(z)$ derived from the best fit model using the combined dataset of ALL16, and the directly observed $D_A(z)$ and $H(z)$ from $H_0$, JLA (the JLA sample of SNe), OHD, gBAO-$9z$ (gBAO measurements at nine effective redshifts) and Ly$\alpha$FB respectively (see {\bf Methods} for detailed explanation and references for data used). The \tc{cyan} horizontal bars indicate the 68\% confidence level (CL) range of Surprise in $\Lambda$CDM, while the dark blue bars correspond to $w(z)$CDM;
Panel b: The corresponding values of Tension $T$, defined as Surprise divided by its standard deviation, shown using the same colour scheme as in Panel a.

\clearpage

\begin{figure}
\begin{center}
\includegraphics[width=0.85\linewidth]{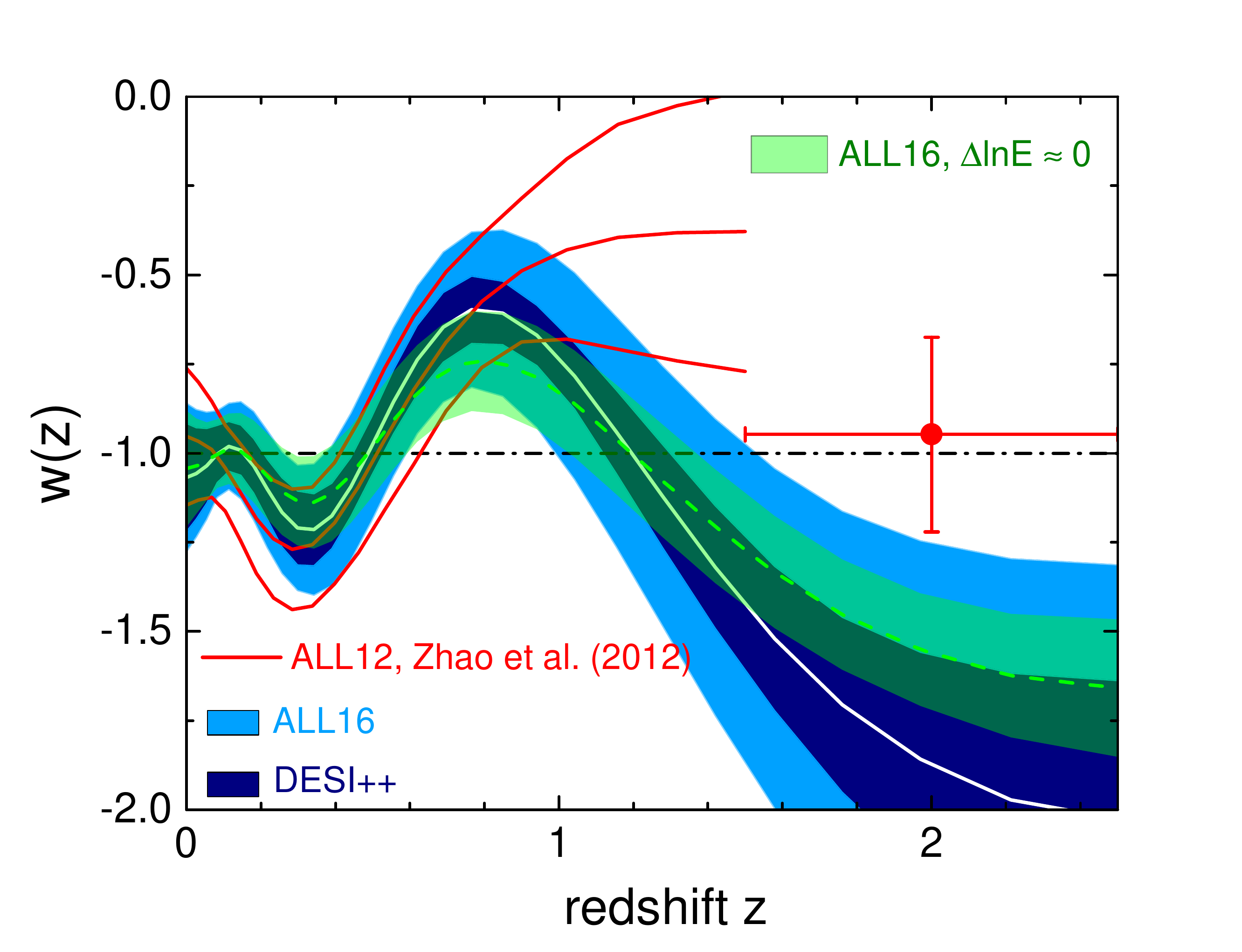}
\end{center}
\end{figure} 

\noindent {\bf Figure 2. The reconstructed evolution history of the dark energy equation of state compared with the 2012 result and the forecasted uncertainty from future data.}\\
The mean (white solid) and the 68\% confidence level (CL) uncertainty (light blue band) of the $w(z)$ reconstructed from ALL16 compared to the ALL12 $w(z)$ reconstructed in Zhao \etal~(2012)\cite{Zhao12} (red lines showing the mean and the 68\% CL band). The red point with 68\% CL error bars is the value of $w(z)$ at $z=2$ ``predicted'' by the ALL12 reconstruction. The dark blue band around the ALL16 reconstruction is the forecasted 68\% CL uncertainty from DESI++. \tc{The green dashed curve and the light green band show the mean and the 68\% CL of $w(z)$ reconstructed from ALL16 using a different prior strength ($\sigma_D=0.4$) for which the Bayesian evidence is equal to that of $\Lambda$CDM. See the text for details.}  

\clearpage

\begin{figure}
\begin{center}
\includegraphics[width=0.8\linewidth]{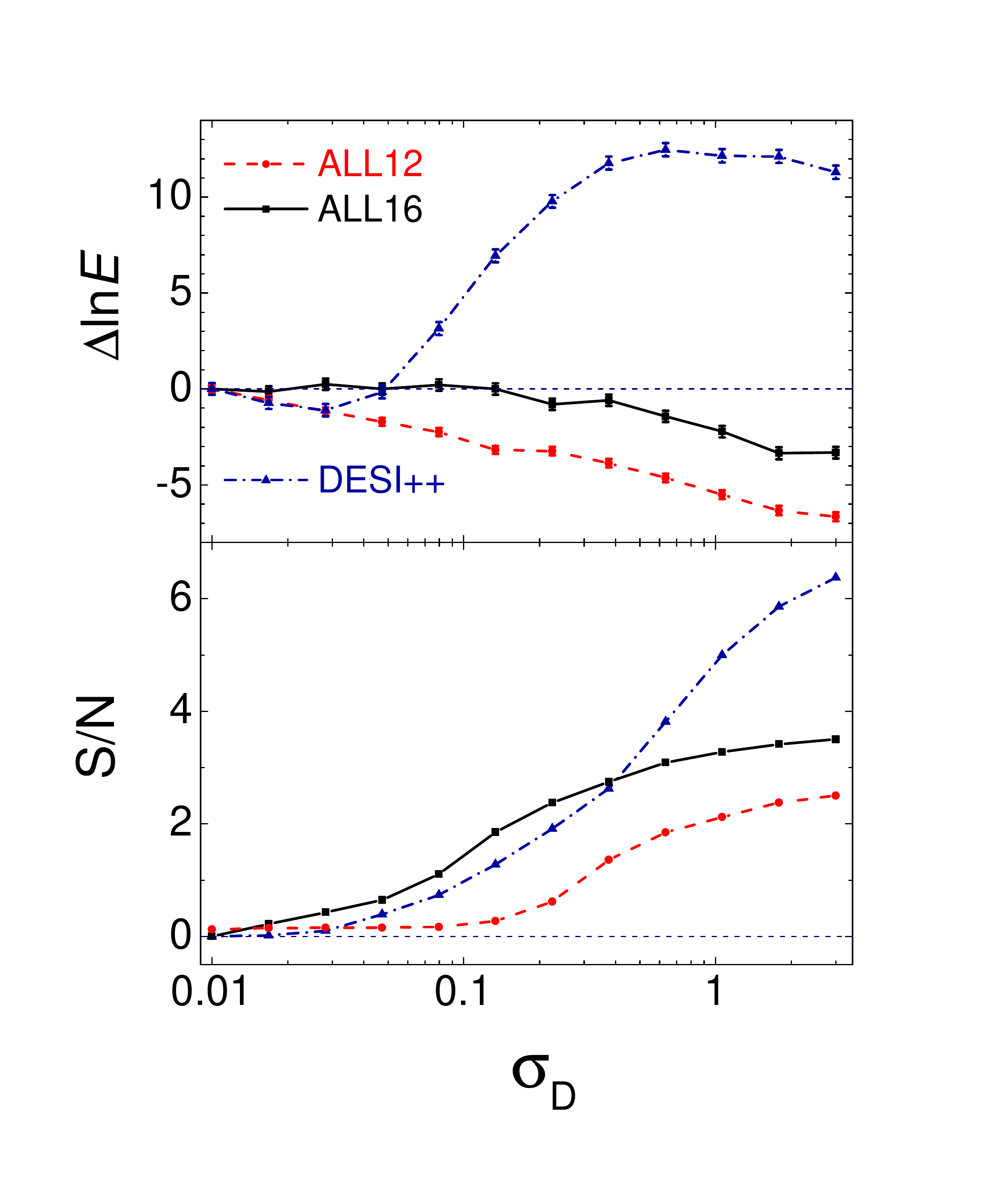}
\end{center}
\end{figure} 

\noindent {\bf Figure 3. The Bayes factor and the significance level of $w\ne-1$ for various correlated priors for current and future data.}\\
The Bayes factor with 68\% confidence level error bars (upper panel) and the statistical significance (lower) of dynamical DE derived from the 2012 data (ALL12; red dashed)\cite{Zhao12}, current data (ALL16; black solid) and future data (DESI++; blue dot-dashed)\cite{DESI} respectively.

\begin{table}
\begin{center}
\begin{tabular}{c|c|c|c|c|c|c|c|c}
\hline\hline
Table 1a									&	P15		&	JLA		&	gBAO		&	$P(k)$	&	WL		&	$H_0$	&	Ly$\alpha$FB		&	OHD		\\
\hline
$\chi^2_{w(z){\rm CDM}}-\chi^2_{\rm \Lambda CDM}$				& $-0.7$		&	$-1.6$	&	$-2.8$			&	$+1.1$	&	$-0.1$	&	$-2.9$	&	$-3.7$			&	$-2.3$	\\
$\chi^2_{w{\rm CDM}}-\chi^2_{\rm \Lambda CDM}$	& $0.0$			&	$+0.5$		&	$+0.7$				&	$+0.4$		&	$+0.2$		&	$-2.9$		&	$-0.2$				&	$0.0$	\\
$\chi^2_{w_0w_a{\rm CDM}}-\chi^2_{\rm \Lambda CDM}$		& $-0.7$		&	$+0.4$	&	$+0.9$			&	$+0.5$	&	$+0.4$	&	$-2.7$	&	$-0.3$			&	$0.0$	\\
\hline\hline
Table 1b			& \multicolumn{3}{c|}{ALL12}		& \multicolumn{3}{c|}{ALL16}		&	\multicolumn{2}{c}{DESI++}	\\
\hline
S/N			&\multicolumn{3}{c|}{$2.5\sigma$}	& \multicolumn{3}{c|}{$3.5\sigma$}	&	\multicolumn{2}{c}{$6.4\sigma$}	\\	
${\rm ln}[E_{w(z){\rm CDM}}/E_{\rm \Lambda CDM}]$	&\multicolumn{3}{c|}{$-6.7\pm0.3$}	& \multicolumn{3}{c|}{$-3.3\pm0.3$}	&	\multicolumn{2}{c}{$~11.3\pm0.3$} \\
\hline\hline
Table 1c	                        &\multicolumn{2}{c|}{$M_1$}  		& \multicolumn{1}{c|}{$M_2$}		 & \multicolumn{2}{c|}{$M_3$} 		&\multicolumn{2}{c|}{$M_4$} 		& \multicolumn{1}{c}{$M_5$}\\
\hline
Definition &\multicolumn{2}{c|}{$M_{\nu}=0.06$ eV}  & \multicolumn{1}{c|}{$M_{\nu}, N_{\rm eff}$} & \multicolumn{2}{c|}{$A_{\rm Lens}$} & \multicolumn{2}{c|}{$w$CDM} & \multicolumn{1}{c}{$w_0w_a$CDM}\\
S/N		       &\multicolumn{2}{c|}{$3.6\sigma$}	& \multicolumn{1}{c|}{$3.4\sigma$}	&	\multicolumn{2}{c|}{$3.4\sigma$} &\multicolumn{2}{c|}{$3.5\sigma$}	& \multicolumn{1}{c}{$3.4\sigma$}	\\
$\Delta {\rm ln}E$                      & \multicolumn{2}{c|}{$-2.8\pm0.3$} & \multicolumn{1}{c|}{$-3.9\pm0.3$} & \multicolumn{2}{c|}{$-3.6\pm0.3$} & \multicolumn{2}{c|}{$-0.9\pm0.3$} & \multicolumn{1}{c}{$0.7\pm0.3$}\\
\hline\hline                        
                       
\end{tabular}
\end{center}
{{\bf Table 1. Statistics of the reconstruction result\\}
\tc{Table 1a: The changes in $\chi^2$ of individual datasets between the ALL16 best-fit $w(z)$CDM, $w$CDM, $w_0w_a$CDM models and the $\Lambda$CDM model; Table 1b: the statistical significance S/N of $w(z)$ deviating from $-1$ based on the improvement in the fit alone (S/N=$\sqrt{\Delta \chi^2}$), and the Bayes factor $\Delta$ln$E$ between the ALL12 and the ALL16 $w(z)$CDM models and the $\Lambda$CDM model, along with the forecast for DESI++; Table 1c: The statistical significance, S/N=$\sqrt{\Delta \chi^2}$, of the preference for $w(z)$CDM over $\Lambda$CDM with added parameters ($M_1,M_2,M_3$), and over a model with a constant $w$ ($M_4$) and one with a linearly varying $w(a)$ ($M_5$). $M_1$ denotes a model with $M_\nu=0.06$ eV, $M_2$ contains varying $M_\nu$ and $N_{\rm eff}$, and $M_3$ includes a varying $A_{\rm Lens}$. For $M_1$, $M_2$ and $M_3$, the new parameters are added to both $w(z)$CDM and $\Lambda$CDM. 
The last row of Table 1c shows the Bayes factors, $\Delta \ln E$, between $w(z)$CDM and the five extended models.}}
\end{table}

\clearpage

\begin{methods}

\subsection{Tension calculation}

The Kullback-Leibler (KL) divergence\cite{Kullback:1951va}, also known as relative entropy, has been extensively utilised as a way of quantifying the degree of tension between different datasets within the $\Lambda$CDM model\cite{Kunz:2006mc, Paykari:2012ne, Amara:2013swa, Seehars:2014ora, Verde:2014qea, Grandis:2015qaa, Seehars:2015qza,Raveri:2016xof}. Rather than focusing on particular model parameters, it is designed to compare the overall concordance of datasets within a given model. \tc{Alternative methods of quantification of the tension have been discussed in the literature\cite{Charnock:2017vcd,Raveri15}.}

The  KL divergence quantifies the proximity of two probability density functions (PDFs), $P_1$ and $P_2$, of a multi-dimensional random variable $\theta$. If both $P_1$ and $P_2$ are assumed to be Gaussian\cite{Seehars:2014ora}, and data are assumed to be more informative than the priors, we can write the difference between the actual and the expected KL divergence, called the ``Surprise''\cite{Seehars:2014ora}, as
\be
\label{eq:surprise}
S=\frac{1}{2{\rm ln}2}\left[(\theta_1 - \theta_2)^T  {\bf {\cal{ C}}}_{1}^{-1}(\theta_1 - \theta_2) -{\bf Tr}\left( {\bf {\cal{ C}}}_{2} \,  {\bf {\cal{ C}}}_{1}^{-1}+\mathbb{I}\right)\right] 
\ee
where $\theta_1$ and $\theta_2$ are the best-fit parameter vectors, ${\bf {\cal{ C}}}_{1}$ and ${\bf {\cal{ C}}}_{2}$ are the covariance matrices for $P_1$ and $P_2$, and $\mathbb{I}$ is the unity matrix. The standard deviation of the expected KL divergence is
\be
\label{eq:sigma}
\Sigma=\frac{1}{\sqrt{2}{\rm ln}2}\sqrt{{\bf Tr}\left( {\bf {\cal{ C}}}_{2} \,  {\bf {\cal{ C}}}_{1}^{-1}+\mathbb{I}\right)^2} \ .
\ee
We can quantify the tension between $P_1$ and $P_2$ in terms of the signal-to-noise ratio $T =S/\Sigma$. If $T\lesssim1$, then $P_1$ and $P_2$ are consistent with each other\cite{Seehars:2015qza}.

\subsection{Datasets used}

The datasets we consider include the Planck 2015 (P15) CMB temperature and polarization auto- and cross-angular power spectra\cite{P15}, the JLA supernovae\cite{JLA} (JLA); the 6dFRS (6dF)\cite{6df} and SDSS main galaxy sample (MGS)\cite{MGS} BAO measurements, the WiggleZ galaxy power spectra\cite{wigglez} in four redshift slices, containing information about the Baryon Acoustic Oscillations (BAO) and Redshift Space Distortions (RSD) ($P(k)$), the weak lensing shear angular power spectra in six redshift slices from CFHTLenS\cite{Heymans:2013fya} (WL), the recent estimate of the Hubble constant $H_0$ obtained from local measurements of Cepheids\cite{Riess16} ($H_0$), the $H(z)$ measurement using the relative age of old and passively evolving galaxies following a cosmic chronometer approach\cite{OHD16} (OHD), the BOSS DR12 ``Consensus" BAO measurement (BAO-$3z$)\cite{Acacia}, the BAO and RSD measurement using the complete BOSS DR12 sample covering the redshift range of $0.2<z<0.75$ at three effective redshifts, the BAO measurement using the same galaxy sample but at nine effective redshifts\cite{Zhao17} (BAO-$9z$), and the Ly$\alpha$ BAO (Ly$\alpha$FB) measurements\cite{Delubac14}. A summary of datasets and data combinations used in this work is shown in Supplementary Table 1. 

\tc{We account for the systematic effects in our analysis as implemented in the public likelihood codes. However, we note that there may be additional systematic effects. For example, the relative velocity between baryons and dark matter may affect the BAO distance measurements\cite{Dalal10}. This effect is estimated to be at sub-percent level for the galaxy BAO measurements of BOSS\cite{BSV,S3pt}, and is currently unknown for Ly$\alpha$FB. For SNe, we use the conventional $\chi^2$ statistics for the analysis, although alternative statistics may extract more information and reduce the systematic effects for the JLA sample to some extent\cite{SNsys,SCF11}.}

\subsection{Non-parametric $w(z)$ reconstruction}

To start, $w(z)$ is parameterised in terms of its values at discrete steps in $z$, or the scale factor $a$. Fitting a large number of uncorrelated bins would result in extremely large uncertainties and, in fact, would prevent the Monte Carlo Markov Chains (MCMC) from converging because of the many degenerate directions in parameter space. On the other hand, fitting only a few bins could significantly bias the result. Our approach is to introduce a prior covariance between the bins based on a specified two-point function that correlates values of $w$ at different $a$, $\xi_w (|a - a'|) \equiv \left\langle [w(a) - w^{\rm fid}(a)][w(a') - w^{\rm fid}(a')] \right\rangle$, which can be taken to be of the form proposed in\cite{Crittenden:2005wj}, $\xi_{\rm CPZ}(\delta a) =  \xi_w (0) /[1 + (\delta a/a_c)^2]$, where $a_c$ describes the typical smoothing scale, and $\xi_w(0)$ is a normalisation factor set by the expected variance in the mean $w$, $\sigma^2_{\bar{w}}$. As shown in\cite{Crittenden11}, results are largely independent of the choice of the correlation function. The prior covariance matrix ${\bf C}$ is obtained by projecting $\xi_w (|a - a'|)$ onto the discrete $w$ bins\cite{Crittenden:2005wj,Crittenden11}, and the prior PDF is taken to be of Gaussian form: ${\cal P}_{\rm prior}({\bf w}) \propto \exp[-({\bf w}-{\bf w}^{\rm fid})^T{\bf C}^{-1}({\bf w}-{\bf w}^{\rm fid})/2]$, where ${\bf w}^{\rm fid}$ is the fiducial model. The reconstructed model is that which maximises the posterior probability, which by Bayes' theorem is proportional to the likelihood of the data times the prior probability,  ${\cal P}({\bf w}|{\bf D}) \propto {\cal P}({\bf D}|{\bf w}) \times {\cal P}_{\rm prior}({\bf w})$. Effectively, the prior results in a new contribution to the total $\chi^2$ of a model, which penalises models that are less smooth. 

In our reconstruction of $w(z)$, we set $a_c=0.06$ and $\sigma_{\bar{w}}=0.04$, which is the ``weak prior" used in Zhao \etal~(2012)\cite{Zhao12}. To calculate the observables, we use a version of ${\tt CAMB}$\cite{CAMB} modified to include DE perturbations for an arbitrary ${\bf w}$\cite{DEP}. We use ${\tt PolyChord}$\cite{polychord}, a nested sampling plug-in for ${\tt CosmoMC}$\cite{Lewis:2002ah}, to sample the parameter space ${\bf P} \equiv (\omega_{b}, \omega_{c}, \Theta_{s}, \tau, n_s, A_s, w_1, ..., w_{30},\mathcal{N})$ where $\omega_{b}$ and $\omega_{c}$ are the baryon and CDM densities, $\Theta_{s}$ is the angular size of the sound horizon at decoupling, $\tau$ is the optical depth, $n_s$ and $A_s$ are the spectral index and the amplitude of the primordial power spectrum, and $w_1,...,w_{30}$ denote the 30 $w$-bin parameters. The first 29 $w$ bins are uniform in $a\in[0.286,1]$, corresponding to $z\in[0,2.5]$, and the last wide bin covers $z\in[2.5,1100]$. We marginalise over nuisance parameters such as the intrinsic SN luminosity.

\subsection{Principal component analysis of $w(z)$}

First, we diagonalise the posterior covariance of $w$-bins to find their uncorrelated linear combinations (eigenmodes), along with the eigenvalues, which quantify how well a given eigenmode is constrained\cite{PCA}. We plot the inverse eigenvalues of the posterior covariance, ordered according to the number of nodes in the eignemodes, in Panel a of Supplementary Fig.~3. The number of nodes is representative of the smoothness in the evolution of eigenmodes, with the first four posterior eigenmodes shown in Panel b of Supplementary Fig.~3. Next, we perform a PCA of the prior covariance matrix and plot its inverse eigenvalues alongside those of the posterior. We see that the fifth and higher number eigenvalues of the two matrices coincide, which means that they are fully determined by the prior. However, the first four inverse eigenvalues of the posterior are significantly larger than that of the prior, indicating that they are constrained primarily by the data. This is precisely the intent of the correlated prior method: the smooth features in $w(z)$ are constrained by the data, with no bias induced by the prior, while the high frequency features are constrained by the prior. Thus, our $w(z)$CDM model effectively has only four additional degrees of freedom compared to $\Lambda$CDM.

\subsection{Dependence of the result on the correlated prior}

To investigate the dependence of our result on the strength of the correlated prior, in Fig.~3, we plot the Bayes factor and the statistical significance of $w\ne-1$ as a function of $\sigma_{\rm D}$, which is a parameter that is added to the inverse covariance matrix of the correlated prior to effectively strengthen it. We find that neither ALL12 nor ALL16 dataset provides evidence for a dynamical DE at all prior strengths, although the Bayes factor for ALL16 is generally much less negative than that of ALL12 for all priors, \eg, it grows from $-6.7\pm0.3$ to $-3.3\pm0.3$ for the prior used in this work, which corresponds to $\sigma_{\rm D}=3$.  On the other hand, the plot shows that, if the $w(z)$CDM model was true, DESI++ would be able to provide a decisive Bayesian evidence over a wide range of prior strengths.

\end{methods}

\begin{addendum}
 \item[Data Availability] The data that support the plots within this paper and other findings of this study are available from the corresponding author upon reasonable request.
\end{addendum}

\clearpage

{\bf \Large Supplementary Information}

{\bf This section contains three supplementary tables and five supplementary figures.}

\begin{table}
\begin{center}
\begin{tabular}{ccc}
\hline\hline
Acronym			&	Meaning							&	References \\
\hline
P15				&	The {\it Planck} 2015 CMB power spectra	&	\cite{si:P15} \\
JLA				&	The JLA supernovae					&	\cite{si:JLA}\\
6dF				&	The 6dFRS (6dF) BAO 				&	\cite{si:6df} \\
MGS				&	The SDSS main galaxy sample BAO 	&	\cite{si:MGS}\\
$P(k)$			&	The WiggleZ galaxy power spectra		&	\cite{si:wigglez}\\
WL				&	The CFHTLenS weak lensing 			&	\cite{si:Heymans:2013fya}\\
$H_0$			&	The Hubble constant measurement		&	\cite{si:Riess16}\\
OHD				&	$H(z)$ from galaxy age measurements 	&	\cite{si:OHD16}\\
gBAO-$3z$		&	3-bin BAO from BOSS DR12 galaxies	&	\cite{si:Acacia}\\
gBAO-$9z$		&	9-bin BAO from BOSS DR12 galaxies	&	\cite{si:Zhao17}\\
Ly$\alpha$FB		& 	The Ly$\alpha$ forest BAO measurements &	\cite{si:Delubac14}\\
\hline
B				&	\multicolumn{2}{c}{P15 + JLA + 6dF + MGS}				\\
ALL12			&	The combined dataset used in Zhao \etal, (2012)\cite{si:Zhao12}\\
ALL16-$3z$		&	\multicolumn{2}{c}{B+$P(k)$+WL+$H_0$+OHD+gBAO-$3z$+Ly$\alpha$FB}	\\
ALL16			&	\multicolumn{2}{c}{B+$P(k)$+WL+$H_0$+OHD+gBAO-$9z$+Ly$\alpha$FB}	\\
DESI++			&	\multicolumn{2}{c}{P15 + mock DESI BAO\cite{si:DESI} + mock SNe\cite{si:EuclidSN}}\\
\hline\hline
\end{tabular}
\end{center}
{\bf Supplementary Table 1. The datasets used in this work.}\\
\end{table}

\clearpage

\begin{table}
\begin{center}
\begin{tabular}{c|c|c}
\hline\hline
			&	ALL12							&	ALL16 \\
	\hline	
SNe			&	SNLS3 (472)~\cite{si:snls3}				&	JLA (740)~\cite{si:JLA}\\
gBAO		&	BOSS DR9; $z_{\rm eff}=0.57$~\cite{si:dr9}	&	BOSS DR12; nine $z_{\rm eff}\in[0.2,0.75]$~\cite{si:Zhao17}\\
Ly$\alpha$FB	&	none								&	BOSS DR11; $z_{\rm eff}=2.34$~\cite{si:Delubac14}\\
$H_0$ [km s$^{-1}$ Mpc$^{-1}$]		&	$74.2\pm3.6$~\cite{si:Riess09}			&	$73.24\pm1.74$~\cite{si:Riess16}\\
CMB			&	WMAP7~\cite{si:wmap7}				&	{\it Planck} 2015~\cite{si:P15} \\
\hline\hline
\end{tabular}
\end{center}
{\bf Supplementary Table 2. Comparing ALL12 and ALL16 datasets.}\\
\end{table}

\begin{table}
\begin{center}
\begin{tabular}{c|ccc|ccc}
\hline\hline
					&						&	$\Lambda{\rm CDM}$	&		&	&	$w(z){\rm CDM}$	&		\\
				\hline
$10^3\Omega_bh^2$	&$22.4\pm0.13$			&$22.5\pm0.16$ 		&$22.4\pm0.14$	 &$22.3\pm0.15$		&$22.4\pm0.21$		&$22.5\pm0.17$ 	\\
$10^3\Omega_ch^2$	&$118.4\pm0.9$			&$121.9\pm2.4$ 		&$118.4\pm1.0$		&$119.2\pm1.4$		&$120.4\pm2.6$		&$117.7\pm1.5$		\\
$10^3\Theta_s$			&$1041\pm0.3$			&$1041\pm0.4$		&$1041\pm0.3$		&$1041\pm0.3$		&$1041\pm0.4$		&$1041\pm0.3$		\\
$10^2\tau$			&$7.9\pm1.6$				&$8.2\pm1.7$			&$4.4\pm1.8$		&$6.9\pm1.7$		&$7.3\pm1.8$		&$4.5\pm1.9$		\\
${\rm ln}(10^{10}A_s)$	&$3.1\pm0.03$				&$3.1\pm0.04$			&$3.0\pm0.04$		&$3.1\pm0.03$		&$3.1\pm0.04$		&$3.0\pm0.04$		\\
$10^2n_s$			&$96.8\pm0.4$				&$97.5\pm0.6$			&$96.8\pm0.4$		&$96.6\pm0.5$		&$96.9\pm0.8$		&$97.0\pm0.5$		\\
$M_{\nu}~({\rm eV})$	&$0.06$					&$<0.19$				&$0$		&$0.06$		&$<0.25$		&$0$		\\
$N_{\rm eff}$		        &$3.04$					&$3.27\pm0.14$		&$3.04$		&$3.04$		&$3.14\pm0.16$		&$3.04$		\\
$A_{\rm Lens}$		        &$1$						&$1$					&$1.17\pm0.07$		&$1$		&$1$		&$1.17\pm0.07$		\\
\hline
$10^3\Omega_m$		&$307\pm5.3$				&$304\pm5.8$			&$301\pm5.6$			&$289\pm11.2$		&$290\pm11.3$		&$285\pm10.9$		\\
$H_0$	 			&$67.9\pm0.40$			&$69.1\pm0.85$		&$68.5\pm0.44$		&$70.2\pm1.3$		&$70.5\pm1.4$		&$70.2\pm1.3$		\\
\hline\hline
\end{tabular}
\end{center}
{\bf Supplementary Table 3. Constraints on cosmological parameters.}\\
The mean and 68\% CL uncertainty on cosmological parameters in both $\Lambda$CDM (left) and $w(z)$CDM (right) models. For the neutrino mass, the 95\% CL upper limit is quoted instead. The unit of the Hubble parameter $H_0$ is km s$^{-1}$ Mpc$^{-1}$.

\end{table}%

\clearpage

\begin{figure}
\begin{center}
\includegraphics[width=0.99\linewidth]{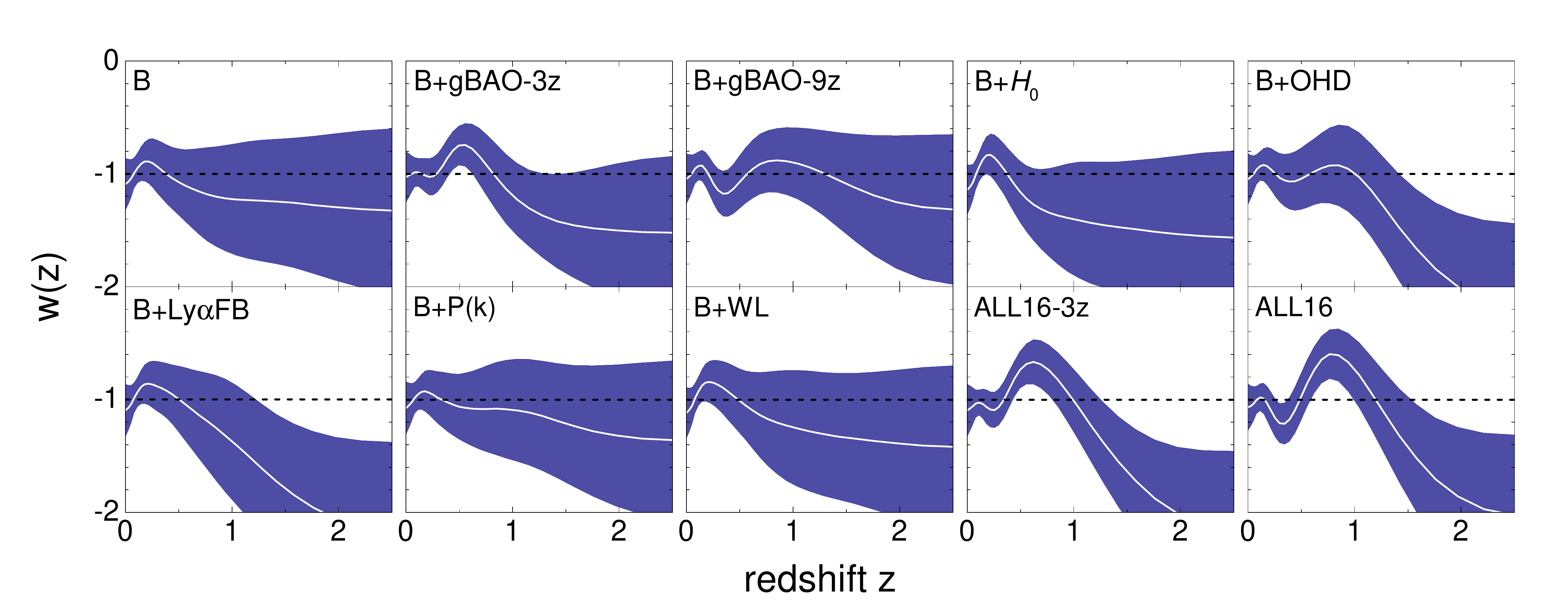}
\end{center}
\end{figure}  

\noindent {\bf Supplementary Figure 1. The reconstructed evolution history of the dark energy equation of state using ten different data combinations.}\\
The reconstructed $w(z)$ (white solid line) and the 68\% CL uncertainty (dark blue shading) from different data combinations shown in the legend. The correlated prior parameters are $a_c=0.06$ and  $\sigma_{\rm m}=0.04$. \\ One can note that the dip in $w(z)$ at $z \sim 0.4$ is more pronounced for ALL16 compared to ALL16-$3z$, thanks to BAO-$9z$ being more informative than BAO-$3z$. As we will see shortly, this makes the ALL16 result more consistent with the $w(z)$ reconstructed in Zhao \etal~(2012)\cite{si:Zhao12} using a different combination of data (ALL 12).\\

\begin{figure}
\begin{center}
\includegraphics[width=0.90\linewidth]{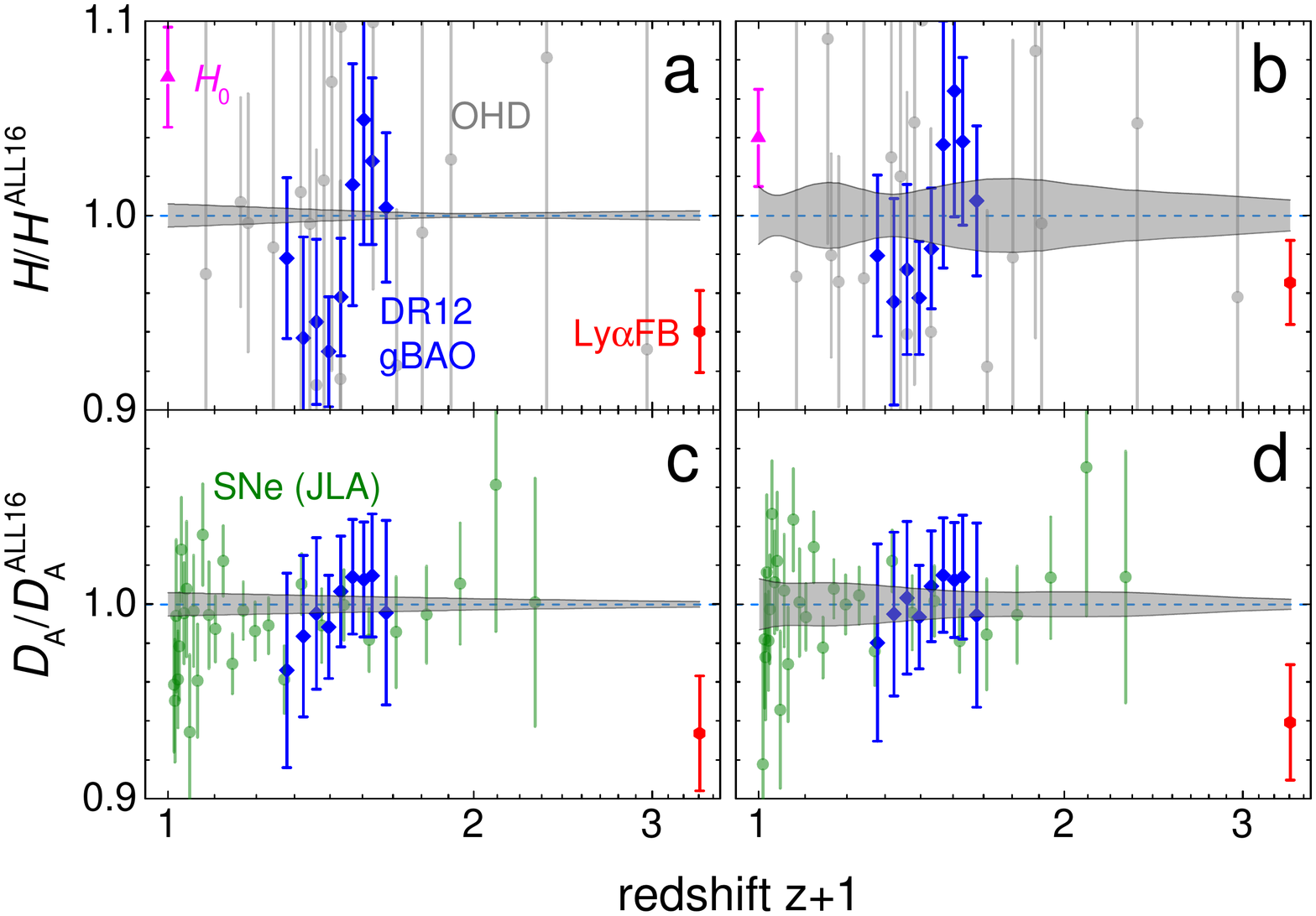}
\end{center}
\end{figure} 

\noindent {\bf Supplementary Figure 2. A comparison between observations and theoretical predictions in $\Lambda$CDM and $w(z)$CDM universes.}\\ The $H(z)$ (panels a, b) and the $D_A(z)$ (c, d) data rescaled by the values derived from the ALL16 best fit $\Lambda$CDM (a, c) and $w(z)$CDM (b, d) models. Datasets are labeled by accordingly coloured text, and the shaded bands indicate the $1\sigma$ uncertainty in the rescaled $H(z)$ and $D_A$. The shaded bands indicate the uncertainty in the rescaled $H(z)$ and $D_A$. One can see that, in the case of $w(z)$CDM, the data points are much more consistent with the corresponding values derived from ALL16, demonstrating the significant reduction in tension.\\

\begin{figure}
\begin{center}
\includegraphics[width=0.99\linewidth]{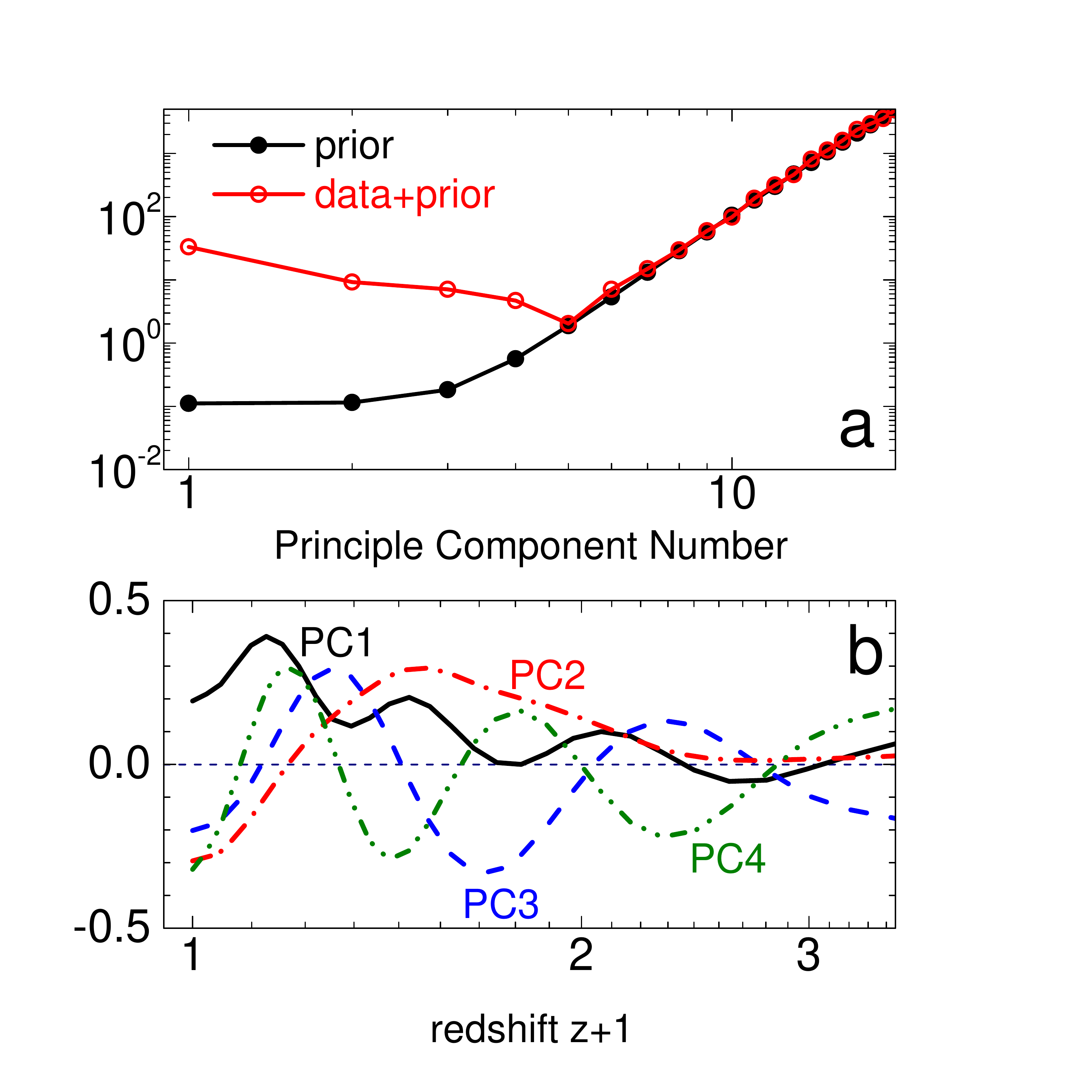}
\end{center}
\end{figure} 

\noindent {\bf Supplementary Figure 3. A principal component analysis of the $w(z)$ reconstruction result.}\\
Panel a: the inverse eigenvalues of the prior covariance matrix (black line with filled dots) and of the posterior covariance (red line with unfilled dots); Panel b: the first four posterior eigenmodes of $w(z)$ for the ALL16 dataset combined with the correlated prior.\\

\begin{figure}
\begin{center}
\includegraphics[width=1\linewidth]{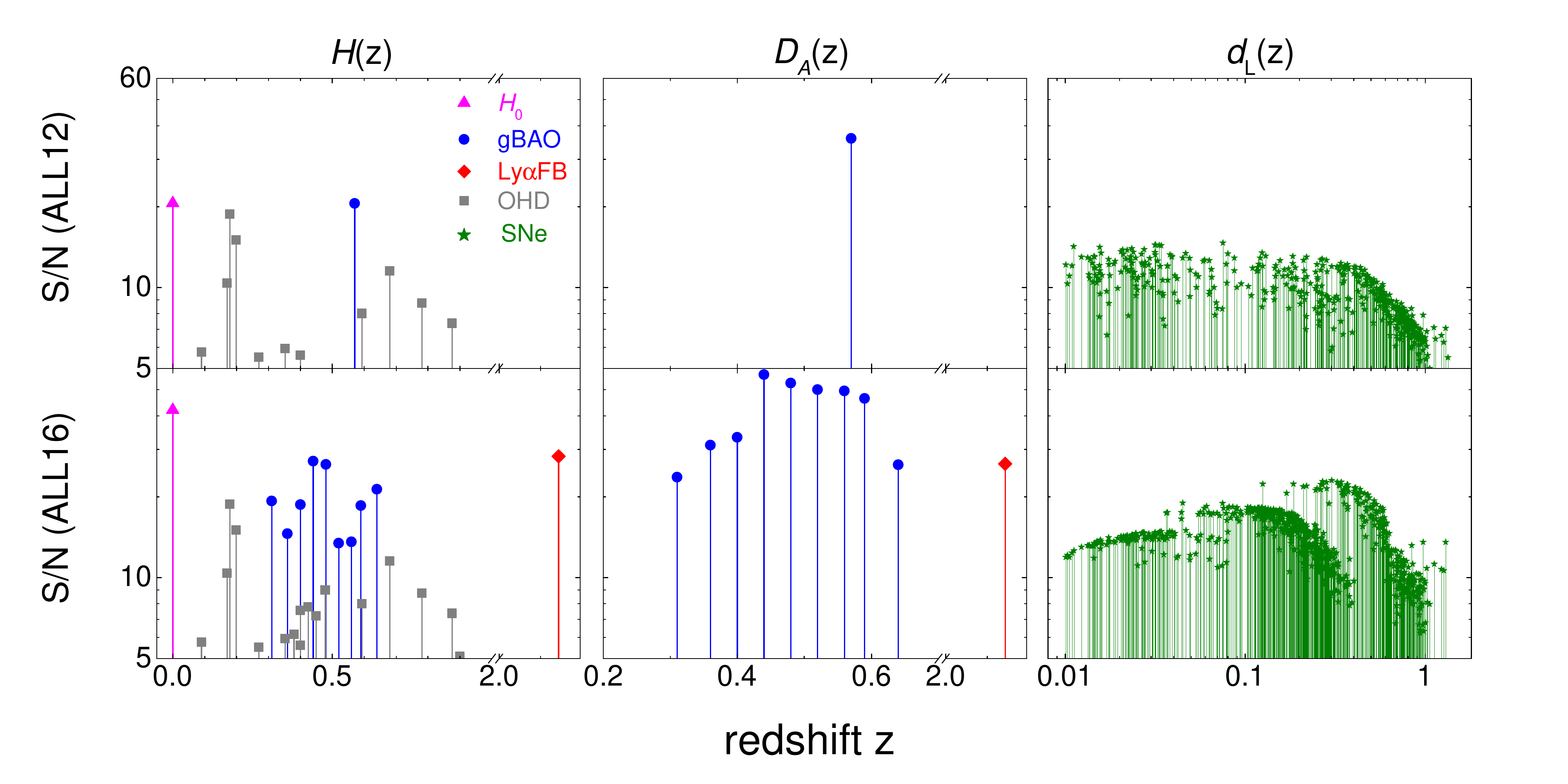}
\end{center}
\end{figure} 

\noindent {\bf Supplementary Figure 4. The signal-to-noise ratio of observables in ALL12 and ALL16 datasets respectively.}\\
The signal-to-noise ratio of the expansion rate of the Universe $H(z)$ (left), angular diameter distance $D_A(z)$ (middle) and luminosity distance $D_L(z)$ for ALL12 (upper) and ALL16 (lower) datasets.

\newpage

\begin{figure}
\begin{center}
\includegraphics[width=1\linewidth]{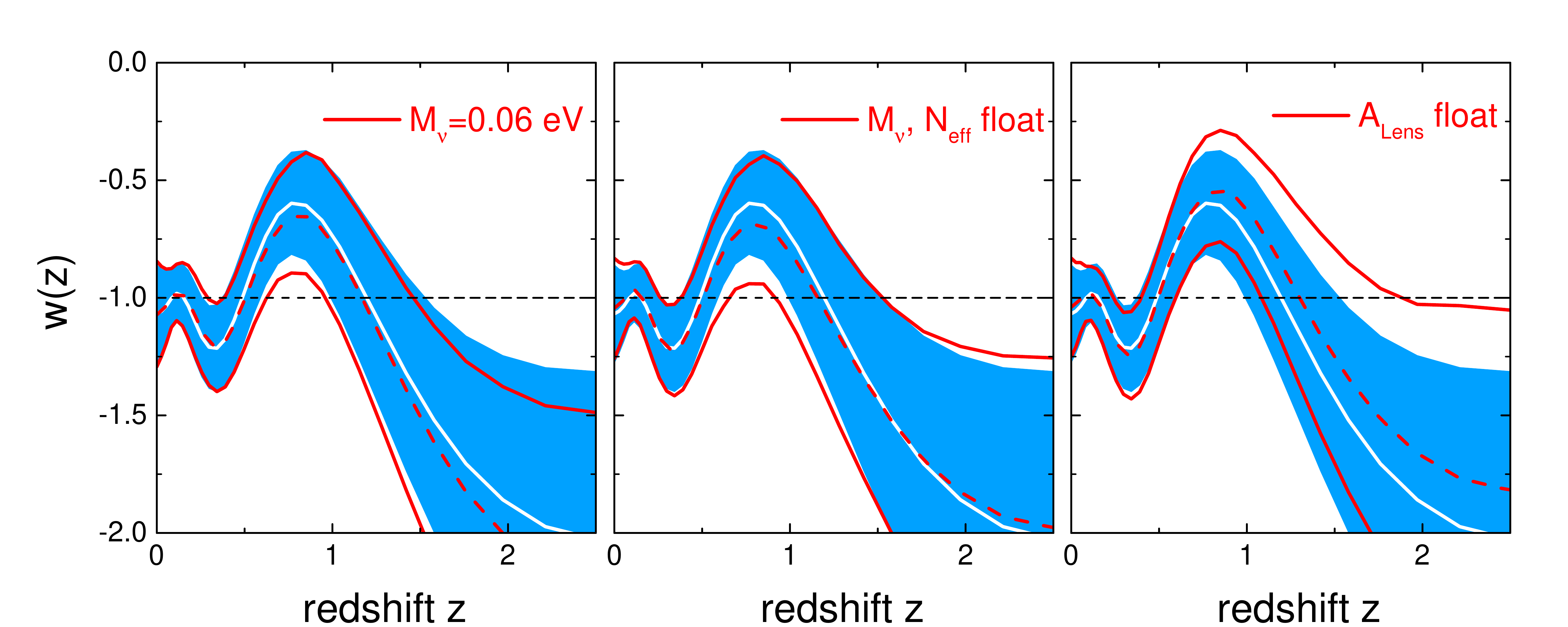}
\end{center}
\end{figure} 

\noindent {\bf Supplementary Figure 5. The effect of neutrinos and CMB lensing amplitude on $w(z)$ reconstruction. }\\
The reconstructed $w(z)$ (red dashed) with 68\% CL uncertainty (red solid) in three cases. Left: the sum of neutrino masses $M_{\nu}$ is fixed to $0.06$ eV; middle: the neutrino masses $M_{\nu}$ and number of relativistic species $N_{\rm eff}$ are marginalised over; right: the CMB lensing amplitude $A_{\rm Lens}$ is marginalised over. The white line and light blue shaded bands in each panel shows the mean and 68\% CL uncertainty of our ALL16 reconstruction, which is shown in Fig. 1 of the Letter using the same colour scheme, for a comparison.

\end{document}